\documentclass[11pt]{article}

\usepackage[letterpaper,top=2cm,bottom=2cm,left=2cm,right=2cm,marginparwidth=2cm]{geometry}
\usepackage{graphicx} 
\usepackage{amsmath, amssymb}
\usepackage[font=small,margin=2cm]{caption}
\usepackage{authblk}
\usepackage{orcidlink}
\usepackage{qcircuit}
\usepackage{siunitx}
\usepackage{multirow}
\usepackage{float}
\usepackage{cite}

\newcommand{\ket}[1]{| #1 \rangle}
\newcommand{\bra}[1]{\langle #1 |}

\newcommand{\ts}{\textsuperscript}

\begin{document}

\title{Hybrid Quantum Physics-Informed Neural Network:\\ Towards Efficient Learning of High-Speed Flows}

\author[]{Fong Yew Leong\,\orcidlink{0000-0002-0064-0118}\thanks{Corresponding author.}}   
\author[]{Wei-Bin Ewe\,\orcidlink{0000-0002-4600-0634}} 
\author[]{Si Bui Quang Tran\,\orcidlink{0000-0002-0888-3931}} 
\author[]{Zhongyuan Zhang\,\orcidlink{0009-0007-5954-7915}} 
\author[]{Jun Yong Khoo\,\orcidlink{0000-0003-0908-3343}} 

\affil[]{\small Institute of High Performance Computing (IHPC), Agency for Science, Technology and Research (A*STAR), 1 Fusionopolis Way, \#16-16 Connexis, Singapore 138632, Singapore}


\maketitle

\begin{abstract}
    This study benchmarks hybrid quantum physics-informed neural network (HQPINN) to model high-speed flows, compared against classical physics-informed neural networks (PINNs) and fully quantum neural networks (QNNs). The HQPINN architecture integrates a parameterized quantum circuit (PQC) with a classical neural network in parallel, trained via a physics-informed loss. Across harmonic, non-harmonic, and transonic benchmarks, HQPINNs demonstrate balanced performance, offering competitive accuracy and stability with reduced parameter cost. Quantum PINNs are highly efficient for harmonic problems achieving the lowest loss with minimal parameters due to their Fourier structure, but struggle to generalize in non-harmonic settings involving shocks and discontinuities. HQPINNs mitigate such artifacts, and with sufficient parameterization, can match the performance of classical models in more complex regimes. Although constrained by current quantum emulation costs and scalability, HQPINNs show promise as general-purpose solvers, offering parameter efficiency with robust fallback behavior, particularly suited for problems where the nature of the solution is not known \emph{a-priori}.
\end{abstract}

\footnotesize
\vspace{1em}
\normalsize

\section{Introduction}

Prediction of compressible air flow features at transonic speed is critical to efficient aerofoil design. This problem involves solving a nonlinear set of partial differential equations (PDEs) in a domain bound by an irregular aerofoil geometry, aspects which pose formidable challenge to traditional computational fluid dynamics (CFD) \cite{Sandeep2021}. In recent years,  machine learning techniques, particularly physics-informed neural networks (PINNs) \cite{Raissi2019}, have emerged as a promising tool for handling high-speed flows, with certain advantages in data-driven neural networks and inverse applications \cite{Mao2020, Jagtap2022, Wassing2024, Cao2024}.

Conceptually, PINNs use differentiable neural networks to approximate the solution of some governing equations, in a way that obeys imposed physical laws, including nonlinear phenomena and geometric constraints \cite{Raissi2019}. This approach allows for the incorporation of prior physical knowledge directly into the learning process, enhancing the model's ability to generalize from limited data. However, as a neural network strategy, PINNs are also limited by extensive use of automatic differentiation for training, which often leads to long training times and prohibitive computational costs \cite{Cao2024}. 

Quantum computing \cite{nielsen2010quantum} offers a promising avenue to accelerate scientific computing and machine learning, particularly through quantum machine learning (QML) \cite{Setty_2025}. QML performs machine learning tasks on a parameterized quantum circuit (PQC), also known as quantum neural network, which exploits quantum features such as superposition and entanglement for higher trainability and expressivity \cite{Abbas2021}. PQCs are not only universal approximators like classical multi-layered perceptrons (MLPs), but they also act as truncated Fourier series models \cite{schuld2021effect}. This unique 
feature enables PQCs to capture complex patterns and relationships in data more efficiently than classical models.

Recent studies have seen efforts to replace classical neural networks with quantum PQC, either partially or entirely \cite{arthur2022,Kordzanganeh2023exponential,Kordzanganeh2023,Jaderberg2023,Jaderberg2023,Sedykh_2024,xiao2024physics,Siegl2024,Trahan2024,Farea2025,Berger2025}. For instance, Arthur and Date \cite{arthur2022} proposed a hybrid quantum neural network consisting of a quantum PQC layer and a classical MLP layer in series for classification problems showing a higher accuracy compared to pure PQC. Kordzanganeh et al. \cite{Kordzanganeh2023exponential} proposed a parallel hybrid quantum neural network capable of extracting both harmonic and non-harmonic features from a dataset \cite{Kordzanganeh2023}. Since then, there are studies of hybrid classical-quantum neural networks modeling flows past a cylinder \cite{Jaderberg2023} and mixing flows via transfer learning \cite{Sedykh_2024}. More recently, Xiao et al. \cite{xiao2024physics} and Siegl et al. \cite{Siegl2024} simulated quantum neural networks for PDEs, which can deliver high accuracy solutions using fewer trainable parameters \cite{Trahan2024,Farea2025}. Berger et al. \cite{Berger2025} leveraged on feedforward neural networks as embedding functions to solve nonlinear PDEs. These studies suggest that hybrid quantum-classical neural network architectures may offer advantages such as enhanced expressivity, in handling more complex PDEs but thus far, work in this area has been limited to problems with smooth solutions \cite{xiao2024physics}.

In this study, by leveraging the benefits of hybrid quantum-classical neural network architectures, we conduct a systematic benchmarking study of HQPINNs on problems ranging from smooth solutions to discontinuous flows and transonic shocks. This is a challenging CFD problem where pure quantum neural networks are \emph{not favored} due to non-harmonic solutions that include shocks and discontinuities  \cite{xiao2024physics,Siegl2024}. To the best of our knowledge, this is the first time HQPINN is applied to high speed flow problems to assess potential of classical and quantum PINNs. We evaluate when hybrid architectures provide benefit, when they fail, and what architectural or training limitations contribute to those outcomes. For verification, we assess the use of HQPINNs on model problems by benchmarking their trainability and training accuracy against classical PINNs and PQCs with varying complexities.

\section{Methodology}
\subsection{Governing equations}
For 2D inviscid compressible flow, the governing Euler equations are \cite{Mao2020, Jagtap2022}:
\begin{align}
    \partial_t U + \nabla \cdot G(U) = 0, \quad \boldsymbol{x} := (x,y) \in \Omega \subset \mathbb{R}^2, \quad t \in (0,T],
    \label{eq:Euler}
\end{align}
where
\begin{align*}
    U = \begin{pmatrix} \rho \\ \rho u \\ \rho v \\ \rho E \end{pmatrix}, \quad 
    G = \{ G_1, G_2 \}, \quad 
    G_1(U) = \begin{pmatrix} \rho u \\ p + \rho u^2 \\ \rho uv \\ pu + \rho uE \end{pmatrix}, \quad 
    G_2(U) = \begin{pmatrix} \rho v \\ \rho uv \\ p + \rho v^2 \\ pv + \rho vE \end{pmatrix},
\end{align*}
$\rho$ is the fluid density, $(u,v)$ are velocity components in $(x,y)$, $E$ is the total energy and $p$ is the pressure. The equation of state for a polytropic gas is
\begin{align}
    p = (\gamma - 1) \left( \rho E - \frac{1}{2} \rho \| \boldsymbol{u} \|^2 \right),
    \label{eq:EOS}
\end{align}
where $\gamma = 1.4$ is the ratio of specific heats for air, and $\boldsymbol{u} = (u,v)$. The pressure $p$ also satisfies the ideal gas law $p = \rho RT$, where T is the temperature and $R$ is the ideal gas constant (287 \si{J.kg^{-1}.K^{-1}} for air). Under steady state conditions, Eq.~\ref{eq:Euler} simplifies to
\begin{align}
    \nabla \cdot G(U) = 0, \quad \{x,y\} \in \Omega.
    \label{eq:Steady Euler}
\end{align}

\subsection{Physics-informed neural networks}

The physics-informed neural network (PINN) consists of an uninformed neural network aiming to satisfy data states including initial and boundary conditions, and an informed neural network aiming to satisfy conservation laws Eq.~\ref{eq:Euler}, both networks sharing hyper-parameters. Let us define 
\begin{align}
    F(\mathbf{x},t) = \partial_t U_{NN}(\mathbf{x},t) + \nabla \cdot G(U_{NN}(\mathbf{x},t)),
    \label{eq:Model}
\end{align}
where $U_{NN}(\mathbf{x},t)$ is the approximation of $U$ by using a deep neural network. The derivative of the deep neural network can be computed using auto differentiation. The parameters of $U_{NN}(\mathbf{x},t)$ can then be learned by minimizing the loss function
\begin{align}
    \mathcal{L} = \mathcal{L}_{BC} + \mathcal{L}_F,
    \label{eq:Loss}
\end{align}
where $\mathcal{L}_{BC}$ is the data loss, including initial and boundary conditions, and $\mathcal{L}_F$ is the physics loss in Eq.~\ref{eq:Model}. Using mean-squared-error (MSE), the respective loss functions can be minimized via
\begin{align}
    \mathcal{L}_{BC} &= \frac{1}{N_{B}}\sum_{j=1}^{N_{B}}
    \left| U_{NN}(\boldsymbol{x}_j^{B},t_j^{B}) -U(\boldsymbol{x}_j^{B},t_j^{B}) 
    \right|^2,
    \label{eq:data_loss}
    \\
    \mathcal{L}_F &= \frac{1}{N_F}\sum_{j=1}^{N_F}
    \left| F(\boldsymbol{x}_j^F,t_j^F)\right|^2,
    \label{eq:physics_loss}
\end{align}
where $\{\boldsymbol{x}_j^{B},t_j^{B}\}_{j=1}^{N_B}$ are initial and boundary collocation points with $N_B$ number of points, and $\{\boldsymbol{x}_j^{F},t_j^{F}\}_{j=1}^{N_F}$ are collocation points for $F(\mathbf{x},t)$ with $N_F$ number of points.

\subsection{Parameterized quantum circuit}

Quantum machine learning (QML) is the application of quantum computing to machine learning tasks~\cite{Benedetti2019}. In the literature, QML circuits are also often referred to as variational quantum circuits, quantum circuit learning, quantum neural networks or parameterized quantum circuits (PQC). 

At its core, a PQC implements a quantum model function $f(\boldsymbol{\theta},\boldsymbol{\varphi})$ by constructing a quantum circuit using fixed gates that encode $N$ data inputs or features $\boldsymbol{\varphi}=(\varphi_1,\dots, \varphi_N)$ and variational gates with $M$ trainable weights $\boldsymbol{\theta}=(\theta_1,\dots, \theta_M)$. Using Dirac notations, the expectation of some observable $\mathcal{O}$ can be measured using the Born rule as
\begin{align}
    f(\boldsymbol{\theta},\boldsymbol{\varphi}) = \bra{\psi(\boldsymbol{\theta},\boldsymbol{\varphi})}\mathcal{O}\ket{\psi(\boldsymbol{\theta},\boldsymbol{\varphi})},
    \label{eq:QML}
\end{align}
where $\ket{\psi(\boldsymbol{\theta},\boldsymbol{\varphi})}$ is the quantum state at the point of measurement. Note that the quantum model fits a truncated Fourier series function~\cite{schuld2021effect}, as 
\begin{align}
    f(\varphi) = \sum_{k=-L}^{L} c_k e^{ik\varphi},
    \label{eq:Truncated Fourier}
\end{align}
where $L$ is the highest degree of the Fourier series. 

Like a multi-layered perceptron (MLP), a PQC is also an universal approximator that can functionally replace at least parts of a neural network, forming a classical-quantum interface known as hybrid quantum neural network (HQNN)~\cite{Kordzanganeh2023}. There are a few reasons why this might be a good idea. Quantum computers already offer exponential computational capacity over classical computers in the form of a large Hilbert space and may also enhance the machine learning process~\cite{Melnikov2023}. Also, the tools of differential quantum circuits~\cite{Kyriienko2021} align with classical neural networks using automatic differentiation and back-propagation. Finally, the Fourier model of a PQC (Eq.~\ref{eq:Truncated Fourier}) could extract harmonic features from a suitable dataset and complement a classical neutral network~\cite{Kordzanganeh2023}.

\subsection{Hybrid quantum physics-informed neural networks}

Here, we propose a hybrid quantum physics-informed neural network (HQPINN) as sketched in Figure~\ref{fig:fig1}. The HQPINN samples $(x, t)$ data and output $\{\rho, u, p\}$ using a hybrid architecture with quantum and classical layers arranged in parallel~\cite{Kordzanganeh2023}: 

\begin{figure}
\centering
\includegraphics[width=0.9\linewidth]{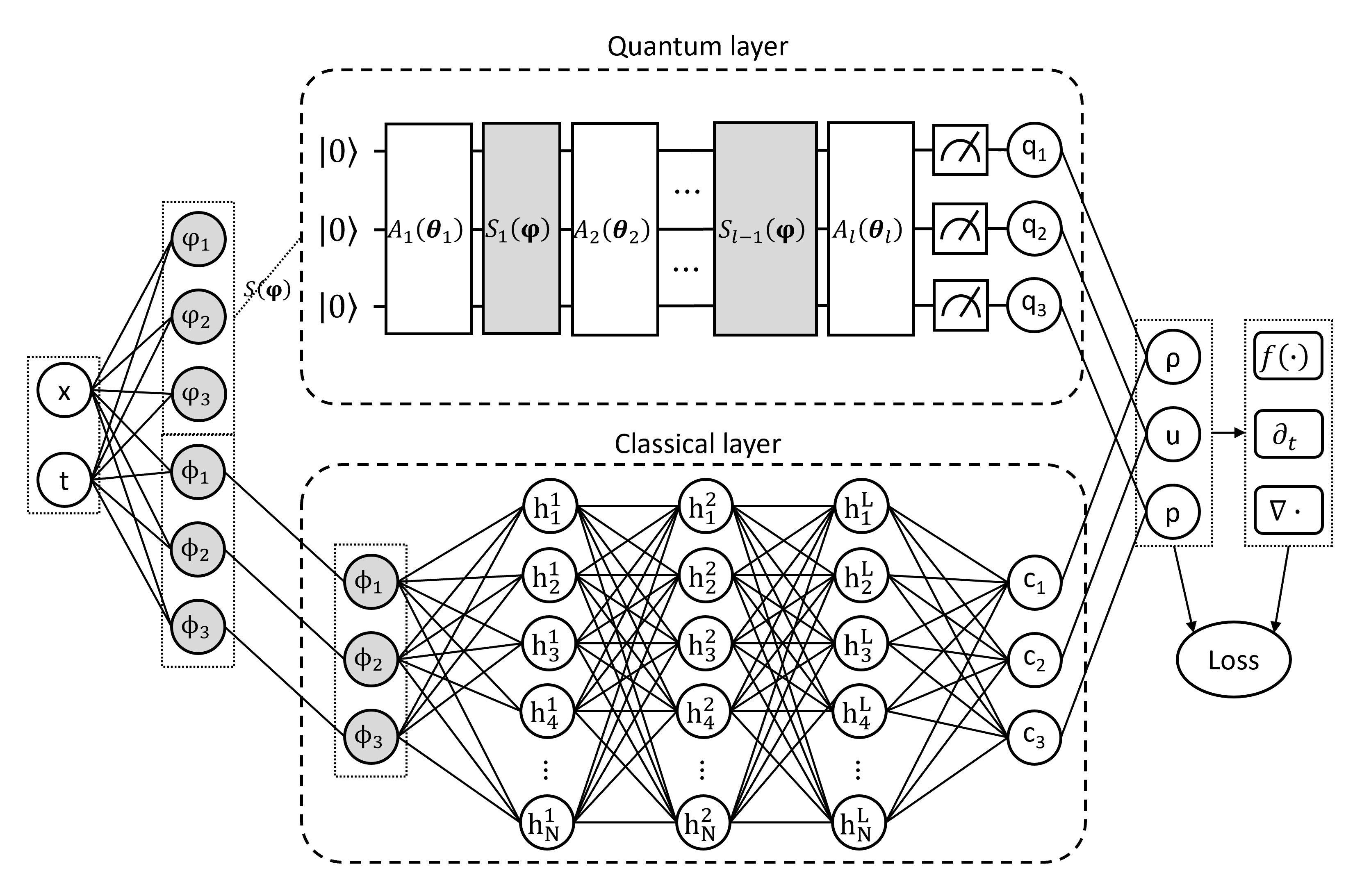}
\caption{Schematic of proposed HQPINN which samples $(x,t)$ data into a parallel architecture coupling a quantum layer (top), an $n$ qubit PQC formed by alternating $l-1$ layers of feature map $S(\boldsymbol{\varphi})$ and $l$ layers of parameterized ansatz $A(\boldsymbol{\theta})$, with a classical layer (bottom), a neural network formed by an input layer with $n$ nodes and $L$ hidden layers of $N$ nodes~\cite{Kordzanganeh2023}. The quantum output $q_{[1,n]}$ and classical output $c_{[1,n]}$ combine linearly to approximate $(\rho,u, p)$ and their informed functions $(f(\cdot),\partial_t, \nabla\cdot)$, both collectively contribute towards a loss function for optimization~\cite{Mao2020}.}
\label{fig:fig1}
\end{figure}

\begin{description}
    \item [Quantum layer] 
    An $n=3$ qubit PQC formed by alternating $l-1$ layers of feature map $S(\boldsymbol{\varphi})$ and parameterized ansatz $A(\boldsymbol{\theta})$ in the following sequence: $A-(S-A)_1- \dots-(S-A)_{l-1}$. Data encoding follows a data-reuploading scheme via angle encoding of the input $n$-node classical layer $\varphi_{1-3} \mapsto S(\boldsymbol{\varphi})$. Refer to Appendix \ref{sec:circuit} for details on quantum circuit design. Pauli-Z expectation measurements are performed on each qubit $q_i = \langle Z_i \rangle$, where $i = [1,n]$, leading to an $n$ node output layer $q_{[1,n]}$. 
    
    \item [Classical layer] Neural network with $(x,t)$ data passed to a $n$-node input layer $\phi_{[1,n]}$ with bias and $\tanh$ activation function applied. This is followed by $L$ hidden layers of $N$ nodes, before a $n$-node output layer $c_{[1,n]}$.
\end{description} 
The final neural network layer is a weighted linear combination of the outputs from the quantum and classical layers $o_i = s_i^c c_i + s_i^q q_i$ \cite{Kordzanganeh2023exponential}, where $s_i^c$ and $s_i^q$ are scalar weights applied to each output component. In this study, we fix $s_i^c = s_i^q = 1$, but future work may consider learnable fusion layers or nonlinear mappings such as $o_i = \mathrm{MLP}([c_i, q_i])$. The combined output is interpreted as predicted physical state $\{\rho, u, p\}$ for minimization of the loss function Eq.~\ref{eq:Loss}. 

We tested the hybrid classical--quantum neural network (Fig.~\ref{fig:fig1}) against a classical--classical neural network where the parallel network layers are both classical layers, and a quantum--quantum neural network where the parallel network layers are both quantum layers. For the introductory 1D harmonic oscillator problem, the HQPINN was found to outperform both classical--classical and quantum--quantum neural network architectures (Appendix~\ref{sec:harmonic_oscillator}).  Henceforth, we aim to verify if the hybrid classical-quantum scheme holds up for more challenging problems such as high-speed shock flows.

\section{Numerical results}
We benchmark classical and quantum PINN models based on two example cases on high-speed flows, one with smooth solutions (\cite{Mao2020}, example 6) and the other with discontinuous solutions (\cite{Mao2020}, example 1). Here we consider three models:
\begin{description}
    \item[Classical--classical(cc)] Parallel classical layers each 4 or 7 hidden layers and 10 or 20 nodes per layer.
    \item[Quantum--quantum(qq)] Parallel quantum layers each 3 qubits with 2 or 4 layers.
    \item[Classical--quantum(hy)] Hybrid parallel classical and quantum layers (see Fig. \ref{fig:fig1}).
\end{description}

To facilitate direct comparison between models, we adopt the following structured notion:
\begin{description}
    \item[cc-\emph{N-L}]: Classical network with \emph{L} layers of \emph{N} nodes.
    \item[qq-\emph{l}]: Quantum network with \emph{l} PQC layers with number of qubits fixed by number of outputs.
    \item[hy-\emph{N-L-l}] Hybrid network with classical layer of size \emph{NL} and quantum layer with $l$ PQC layers.
\end{description}

Models are implemented on \emph{Pytorch} \cite{Paszke2019} interfaced with \emph{Pennylane} \cite{Bergholm2018} and run on distributed CPU without GPU acceleration. Loss functions (Eq. \ref{eq:Loss}) are minimized via mean square error (Eq. \ref{eq:data_loss} and \ref{eq:physics_loss}) using Adam optimizer for 20\,000 steps with learning rate 0.0005. Quantum layers are simulated classically, which imposes a significant time overhead compared to classical layers, by at least two orders of magnitude for the same number of trainable parameters \cite{xiao2024physics,Siegl2024}.

\subsection{Smooth Euler Equation (Harmonic Regime)}\label{sec:smooth}

Consider a 1D Euler equation with smooth solution (\cite{Mao2020}, example 6). We use periodic boundary conditions and initial condition,
\begin{align}
    U_0=(\rho_0, u_0, p_0)=(1.0+0.2\sin(\pi x), 1.0, 1.0),
    \label{eq:ex6_ic}
\end{align}
which leads to a traveling wave solution 
\begin{align}
    (\rho,u,p)=(1.0+0.2\sin(\pi (x-t)), 1.0, 1.0),
    \label{eq:ex6_exact}
\end{align}
in a domain defined by $x\in(-1,1)$ and $t\in(0,2)$. Following (\cite{Mao2020}, example 6), 
we randomly sample $N_{ic}=50$, $N_{bc}=50$ and $N_{F}=2000$ training points for initial condition $ic$, boundary condition $bc$ (Eq. \ref{eq:data_loss}) and domain $F$ (Eq. \ref{eq:physics_loss}) respectively. 

\begin{figure}
\centering
\includegraphics[width=0.7\linewidth]{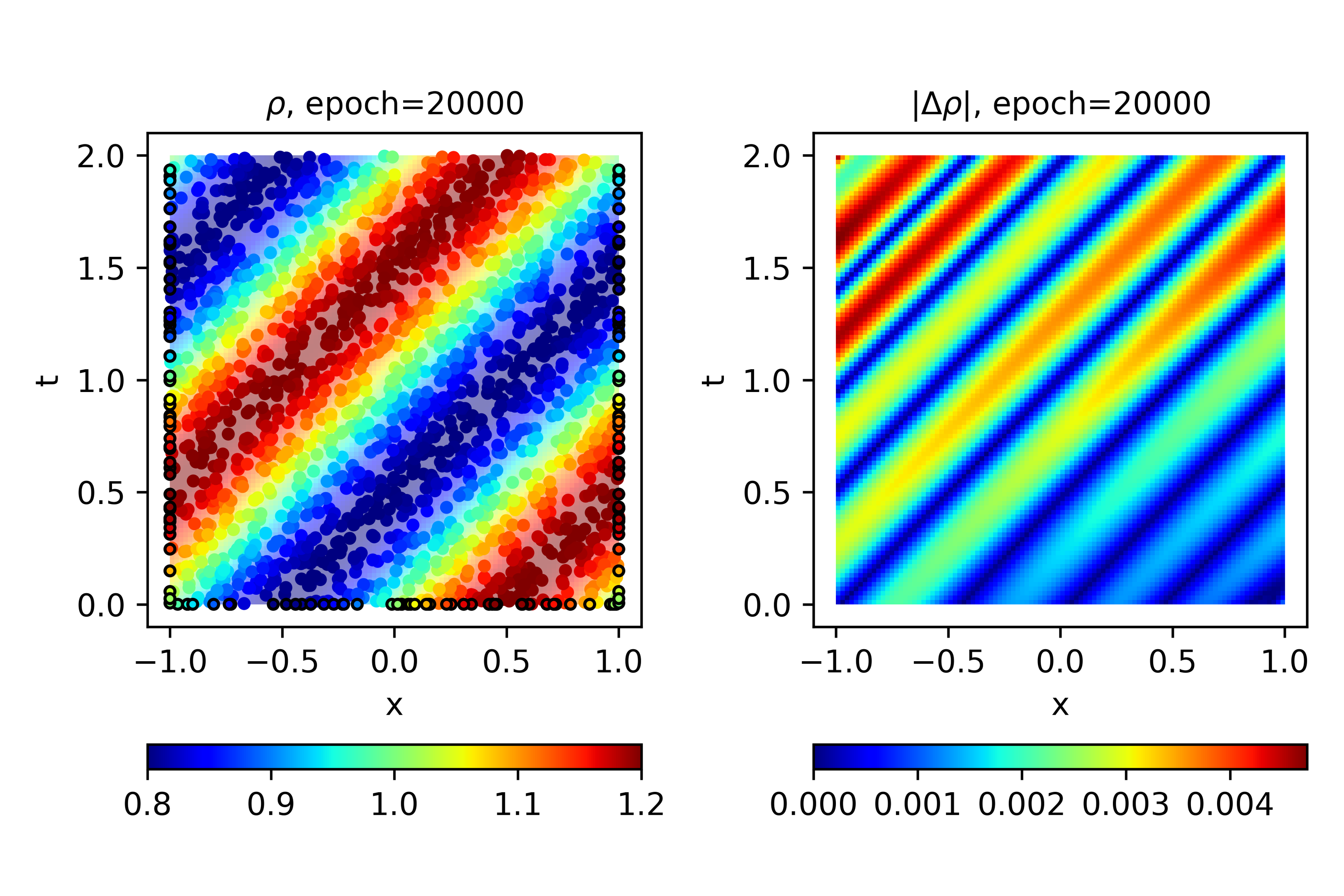}
\caption{Plot of smooth traveling wave solution for 1D Euler equation (Section \ref{sec:smooth}). (Left) Density in x--t domain. Opaque bubbles are domain residual training points and dark circled bubbles are boundary data training points. (Right) Normalized deviation from exact solution (Eq. \ref{eq:ex6_exact}). }
\label{fig:fig2}
\end{figure}

Figure \ref{fig:fig2} shows a traveling wave solution of the density based on classical \emph{cc-10-4} model, with high fidelity agreement with the exact solution indicated by their normalized deviation. Table \ref{tab:smooth} (Appendix \ref{sec:tables}) compares the number of trainable parameters required for each model and the resultant loss and error at the 20\,000\ts{th} epoch. 

Here, the best model is the quantum \emph{qq-2} model with an average loss of $2.63\times10^{-8}$, a tenfold improvement compared to the best classical \emph{cc-10-7} model with an average loss of $3.16\times10^{-7}$. In addition, the quantum \emph{qq-2} model requires only 87 classical and 36 quantum trainable parameters vastly outperforming the best classical \emph{cc-10-7} model. However, note that larger quantum models suffer from poor trainability due to excessive expressibility, a phenomenon also observed in QML elsewhere \cite{xiao2024physics}. 

The hybrid model \emph{hy-10-4-2} achieves strong balanced performance with high accuracy and trainability, as well as reduced parameter count (484 vs. 1505). Interestingly, the best hybrid model \emph{hy-10-4-2} is not a combination of the best classical \emph{cc-10-7} and quantum \emph{qq-2} models.

\begin{figure}[h]
\centering
\includegraphics[width=0.9\linewidth]{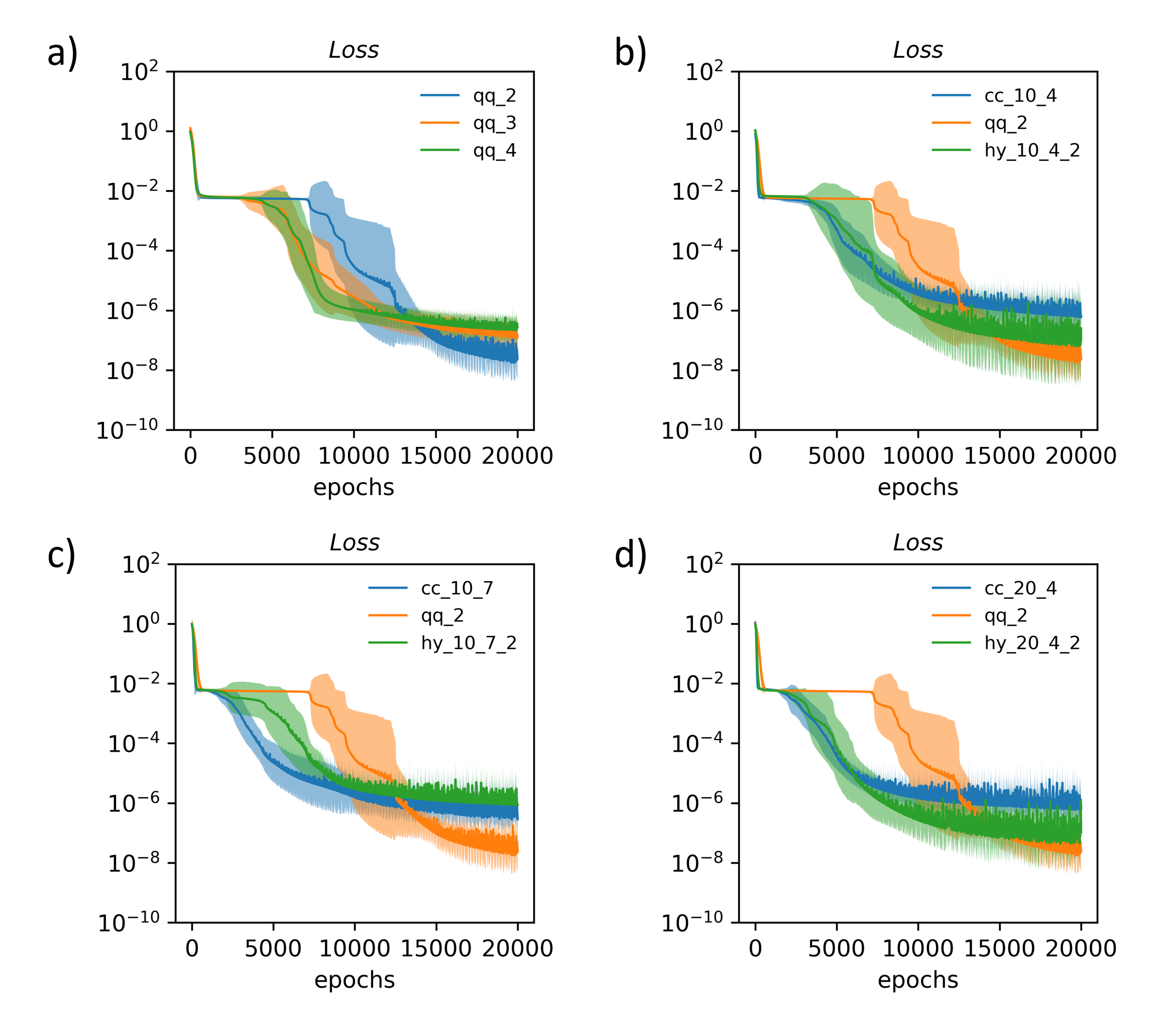}
\caption{Training loss against epochs for smooth PINN solution of 1D Euler equation (Section \ref{sec:smooth}). (a) Larger quantum models improve trainability at the cost of accuracy. Hybrid models outperform both (b) shallow \emph{cc-10-4} and (d) wide \emph{cc-20-4} classical models, but not (b) deep \emph{cc-10-7} classical model.}
\label{fig:fig3}
\end{figure}

Figure \ref{fig:fig3} shows training loss against epochs for selected neural network models with shallow, deep or wide characteristics. Our results show that classical models train fast but with low accuracy, in contrast to quantum models which train slower but with high accuracy. Hybrid models inherit high trainability from the classical layer and superior accuracy from quantum layer, the best features from each model. The exception is the deep \emph{hy-10-7-2} model (Fig. \ref{fig:fig3}c), where training is weighted towards the classical layer than the quantum layer.

\subsection{Discontinuous Euler Equation (Contact Discontinuity)}\label{sec:shock}

Consider a 1D Euler equation with moving contact discontinuity (\cite{Mao2020}, example 1). We use Dirichlet boundary conditions,
\begin{align}
    (\rho_L, u_L, p_L)=(\rho_R, u_R, p_R)=(1.0, 0.1, 1.0),
    \label{eq:ex1_bc}
\end{align}
where subscripts $L$ and $R$ refer to left and right boundaries respectively. The exact solutions are
\begin{equation}
  \rho(x,t) =
    \begin{cases}
      1.4, & x<0.5+0.1t,\\
      1.0, & x>0.5+0.1t,\\
    \end{cases}
    \quad u(x,t)=0.1, \quad p(x,t)=1.0,
    \label{eq:ex1_exact}
\end{equation}
in a domain defined by $x\in(0,1)$ and $t\in(0,2)$. Following (\cite{Mao2020}, example 1), 
we randomly sample $N_{ic}=60$, $N_{bc}=60$ and $N_{F}=1000$ training points for initial condition $ic$, boundary condition $bc$ and domain $F$ respectively. 

\begin{figure}
\centering
\includegraphics[width=0.7\linewidth]{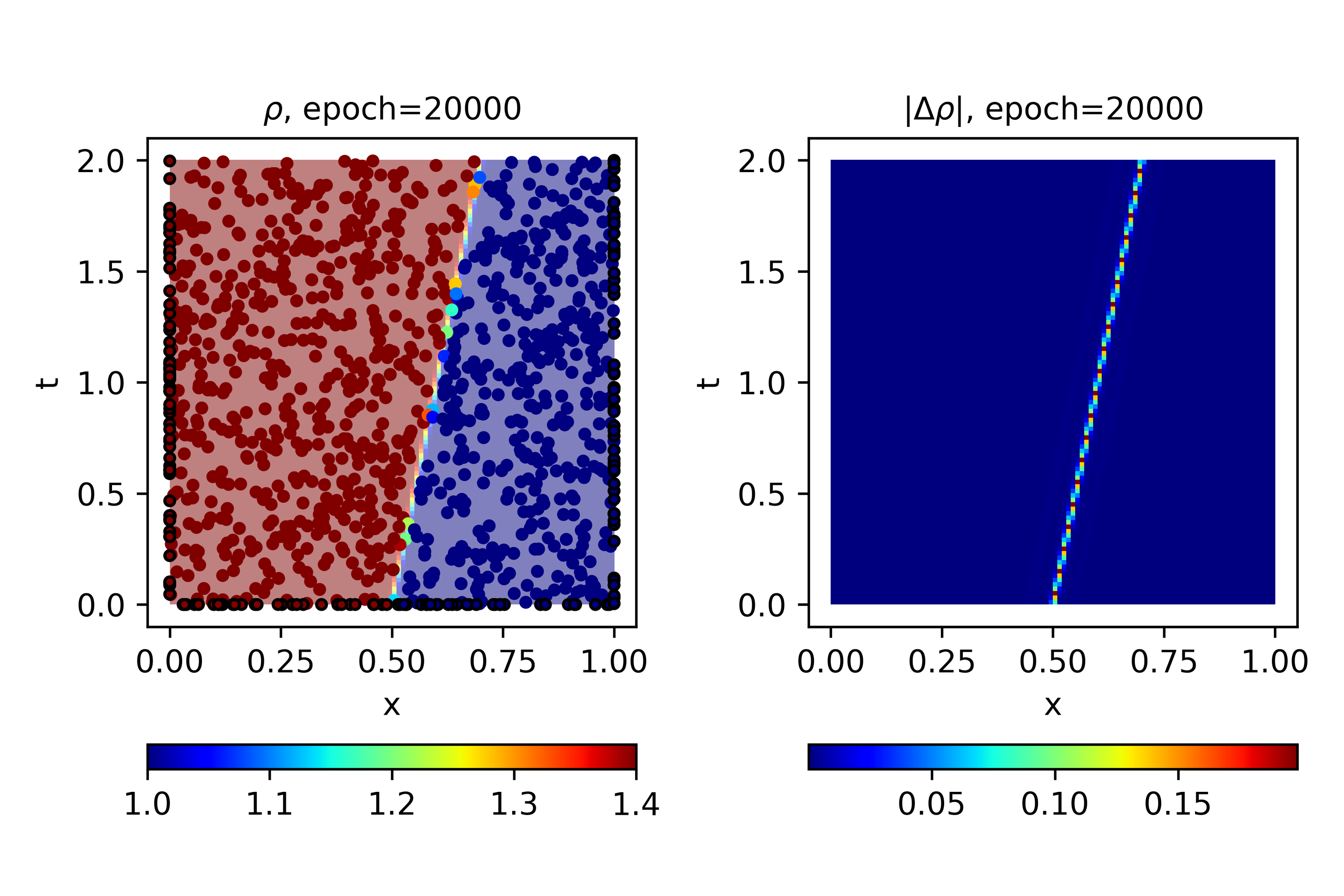}
\caption{Plot of moving contact discontinuity for 1D Euler equation (Section \ref{sec:shock}). (Left) Density in x--t domain. Opaque bubbles are domain residual training points and dark circled bubbles are boundary training points. (Right) Normalized deviation from exact solution (Eq. \ref{eq:ex1_exact}). }
\label{fig:fig4}
\end{figure}

Figure \ref{fig:fig4} shows moving contact discontinuity in the density based on classical \emph{cc-10-4} model, where the normalized deviation from exact solution concentrated along the discontinuity itself. Table \ref{tab:shock} (Appendix \ref{sec:tables}) compares the trainable parameters required for each model and the resultant loss and error at the 20\,000\ts{th} epoch. The shallow classical \emph{cc-10-4} model is the best model with an average loss of $6.89\times10^{-7}$ for 845 trainable parameters. 
In contrast, quantum \emph{qq} models significantly underperformed with average loss $\overline{\mathcal{L}}_{qq}\sim \mathcal{O}(10^{-4})$, but which improves with increasing number of trainable quantum parameters (average loss decreases by 72\% from \emph{qq-2} to \emph{qq-4}). Overall, the hybrid models ($\overline{\mathcal{L}}_{hy} \sim \mathcal{O}(10^{-6}-10^{-5})$) performed slightly worse than classical models ($\overline{\mathcal{L}}_{cc} \sim \mathcal{O}(10^{-7}-10^{-6})$), but significantly better than quantum models ($\overline{\mathcal{L}}_{qq}\sim \mathcal{O}(10^{-4})$). 

\begin{figure}
\centering
\includegraphics[width=0.9\linewidth]{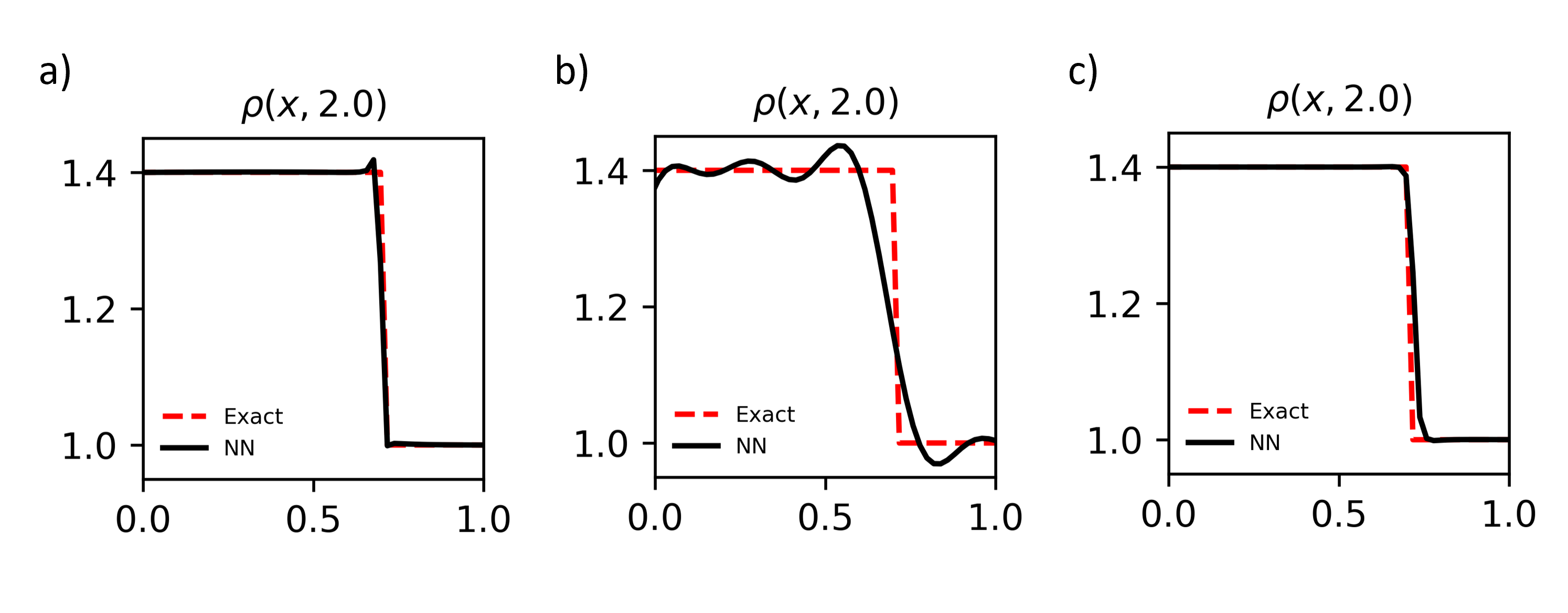}
\caption{Density profiles at time $t=2$ for (a) classical \emph{cc-10-4}, (b) quantum \emph{qq-2} and (c) hybrid \emph{hy-10-4-2} models compared to exact solution (red dashed line). The quantum model (b) attempts to fit a harmonic waveform to the shock problem, leading to overshoot and under-damped oscillations; this is compensated by the hybrid model (c) due to the presence of a classical layer.}
\label{fig:fig5}
\end{figure}

Figure \ref{fig:fig5} shows density profiles at time $t=2$ where we see the quantum model exhibits spurious oscillations and fails to capture the shock discontinuity. In contrast, hybrid models are able to reduce overshoot observed in quantum models and achieve high solution fidelities similar to classical models.

\begin{figure}
\centering
\includegraphics[width=0.9\linewidth]{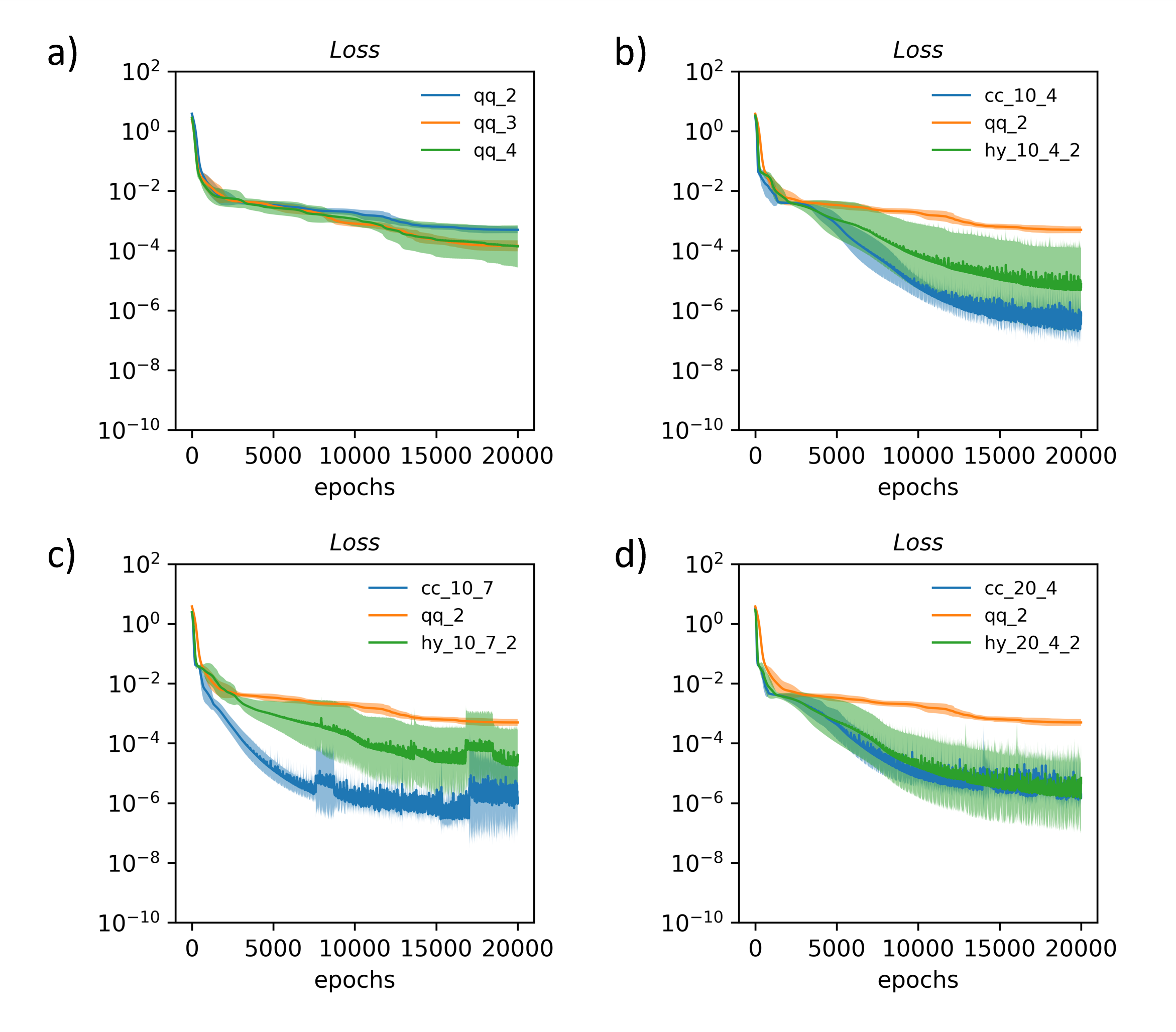}
\caption{Training loss against epochs for discontinuous PINN solution of 1D Euler equation (Section \ref{sec:shock}). (a) Quantum models perform poorly for shock problems. Hybrid models outperform quantum models, but underperform (b) shallow \emph{cc-10-4} and (c) deep \emph{cc-20-7} classical models. (d) The wide hybrid \emph{hy-20-4-2} model is comparable to the classical \emph{cc-20-4} model in both trainability and accuracy.}
\label{fig:fig6}
\end{figure}

Figure \ref{fig:fig6} shows training loss against epochs for selected neural network models with shallow, deep or wide characteristics. We found that the classical models outperformed the quantum models in all cases, while the hybrid models suffer from significant variance in both accuracy (loss $\mathcal{L}$) and trainability (rate of change of $\mathcal{L}$). Figure \ref{fig:fig6}a-c shows that hybrid models are unable to match the respective classical models in terms of loss accuracy. In addition, the deep hybrid \emph{hy-10-7-2} model performs worse than deep classical \emph{cc-hy-10-7} in terms of trainability (Fig. \ref{fig:fig6}c). Importantly, note that the wide hybrid \emph{hy-20-4-2} model performs as well as the wide classical \emph{cc-20-4} model (Fig. \ref{fig:fig6}d), while requiring only about 55\% of the number of trainable parameters. This suggests that hybrid models may be advantageous for problems with non-harmonic solution under over-parameterized conditions.

\subsection{2D Transonic Aerofoil Flow}\label{sec:foil}
We test HQPINN on a 2D steady Euler equation (Eq. \ref{eq:Steady Euler}) for transonic flow past an NACA0012 aerofoil with chord $(0,1)$ set on a square Cartesian grid $x\in(-1,3.5), y\in(-2.25,2.25)$. Setting $U=(\rho, u, v, T)$ for density, horizontal and vertical velocities, and temperature, we apply an inlet boundary condition $U_{in}=(\rho_{in}, u_{in}, v_{in}, T_{in})=(1.225, 272.15, 0.0, 288.15)$, outlet boundary condition $P_{out} = 0$ and periodic boundary conditions on the side boundaries, all in SI units.

The flow equations are solved using Computational Fluid Dynamics (CFD) on finite volume method with mesh refinement around the aerofoil surface and free slip boundary condition applied (refer to Appendix \ref{sec:cfd} for details). This is followed by PINN training, where we randomly sample 40 boundary points along each boundary for data training (Eq. \ref{eq:data_loss}) and 4000 domain points for physics loss training (Eq. \ref{eq:physics_loss}). 

It should be pointed out that the transonic shock problem is, in fact, highly challenging even for vanilla PINNs. It is only until recently that add-on strategies, such as coordinate transformation, artificial viscosity, adaptive weights and shock sensors \cite{wassing2024physics, Ren2024, McClenny2023}, enable PINNs to handle shock regions. While exploring these strategies are beyond the scope of this work, to mitigate the expression of shocks, we apply an adaptive gradient-weight \cite{Liu2024},
\begin{align}
    \lambda=\frac{1}{\varepsilon(|\nabla\cdot\vec u |-\nabla\cdot\vec u )+1},
    \label{eq:lambda}
\end{align}
where $\vec u$ is the velocity and $\varepsilon$ is a parameter (here taken as 0.1). To facilitate training, we randomly assigned 400 domain points to data training instead of PDE training. Loss functions are minimized via mean square error (Eq. \ref{eq:data_loss} and \ref{eq:physics_loss}) using Adam optimizer for 40\,000 steps with learning rate 0.0005, followed by L-BFGS optimizer for 2000 steps. 

\begin{figure}
\centering
\includegraphics[width=1\linewidth]{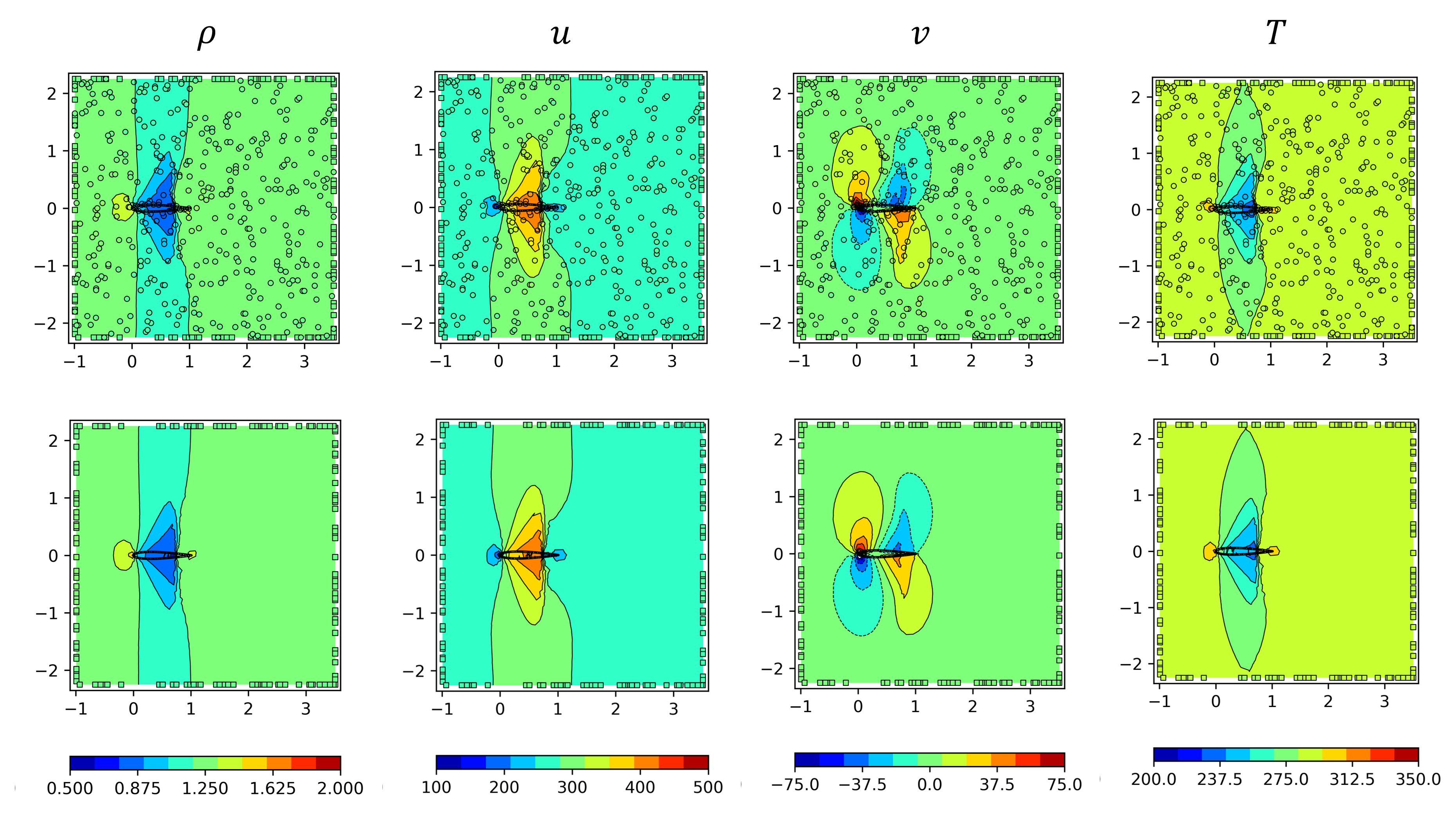}
\caption{Solution plots of density ($\rho$), horizontal and vertical velocities ($u,v$), and temperature ($T$) based on (top) classical \emph{cc-40-4} model (training points in open circles), and (bottom) CFD solution.}
\label{fig:fig7}
\end{figure}

Figure \ref{fig:fig7} compares solutions for density, velocity and temperature for classical \emph{cc-40-4} model against ground solutions obtained from CFD (Appendix \ref{sec:cfd}). Solutions for quantum and hybrid models can be found in Appendix \ref{sec:quantum_hybrid}. Note that the stagnation regions are small and may not resolve properly without sufficient training points sampled from within.

\begin{figure}
\centering
\includegraphics[width=0.9\linewidth]{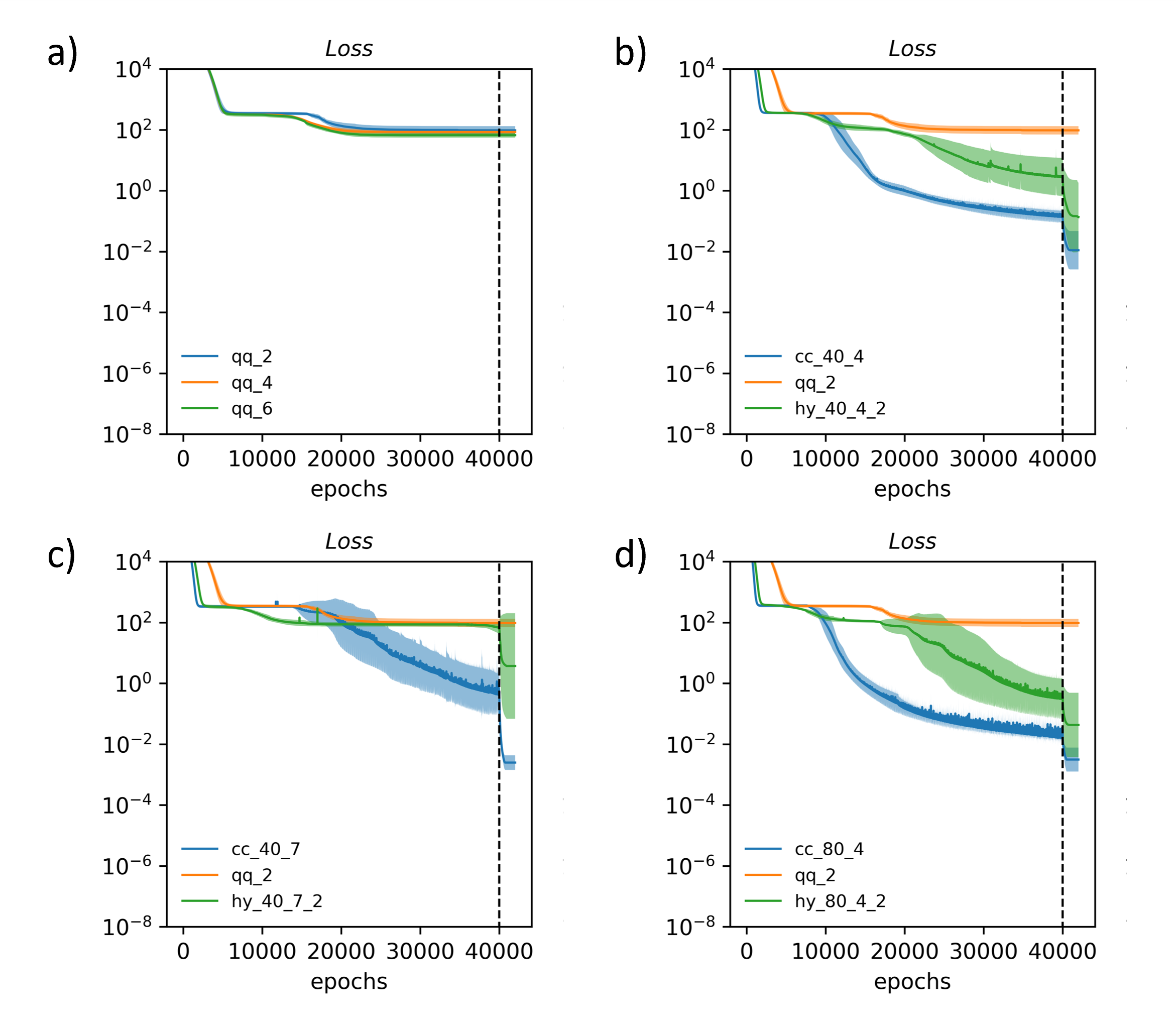}
\caption{Training loss against epochs for PINN solution of 2D transonic aerofoil problem (Section \ref{sec:foil}). (a) Quantum models perform poorly for shock problems. Hybrid models outperform quantum models but underperform (b) shallow \emph{cc-40-4}, (c) deep \emph{cc-40-7} and (d) wide \emph{cc-80-4} classical models.}
\label{fig:fig8}
\end{figure}

Because the transonic flow requires a minimum of 40 nodes per hidden layer to resolve, quantum models are significantly under-parameterized in comparison (Table \ref{tab:foil}, Appendix \ref{sec:tables}). Figure \ref{fig:fig8} shows that under-parameterized quantum models do not scale in trainability and accuracy, which in turn also limits the trainability of hybrid models.

It is currently challenging to study the performance of larger quantum models as the time required to simulate them classically scales unfavorably. This bottleneck may be relieved in the future as fault-tolerant quantum hardware or accelerated hybrid software scheme become available.

\subsection{Summary of Results}\label{sec:insight}

This section highlights key insights based on model performances across the three benchmark problems. Here, we benchmark trainability using the epoch ratio defined as the ratio of number of epochs required to reach a specified threshold loss and total number of simulated epochs; the threshold loss are set as \{Harmonic: $10^{-5}$, Non-harmonic: $10^{-4}$, Transonic: $10$\} for the respective problems. Also, we benchmark costs using the parameter ratio defined as the ratio of number of parameters of the model and that of the classical model of the problem. Table \ref{tab:summary} summarizes loss, epoch ratio, and parameter ratio for representative configurations of classical, quantum, and hybrid PINNs. 

\begin{table}[H]
    \centering

    \begin{tabular}{ccccc}
        \hline
        \multirow{2}{2.5cm}{\centering Harmonic (Sec. \ref{sec:smooth})} & 
        \multirow{2}{2.0cm}{\centering Size} &
        \multirow{2}{2.0cm}{\centering Loss} &
        \multirow{2}{1.5cm}{\centering Epoch ratio} &
        \multirow{2}{2.5cm}{\centering Parameter ratio}
        \\ 
        
        \\
        \hline
        \noalign{\vskip 1mm}
        
        Classical&  10-4& $\sim10^{-7}$&  0.260& 1.000\\
        Quantum&  2& $\sim10^{-8}$&  0.675& 0.146\\
        Hybrid&  10-4-2& $\sim10^{-7}$&  0.350&  0.573\\
        
        \noalign{\vskip 1mm}
        \hline
    \end{tabular}

        \vskip 3mm

    \begin{tabular}{ccccc}
        \hline
        \multirow{2}{2.5cm}{\centering Non-harmonic (Sec. \ref{sec:shock})} & 
        \multirow{2}{2.0cm}{\centering Size} &
        \multirow{2}{2.0cm}{\centering Loss} &
        \multirow{2}{1.5cm}{\centering Epoch ratio} &
        \multirow{2}{2.5cm}{\centering Parameter ratio}
        \\ 
        
        \\
        \hline
        \noalign{\vskip 1mm}
        
        Classical&  20-4& $\sim10^{-7}$&  0.310& 1.000\\
        Quantum&  2& $\sim10^{-4}$&  $>$ 1& 0.056\\
        Hybrid&  20-4-2& $\sim10^{-6}$&  0.380&  0.522\\
        
        \noalign{\vskip 1mm}
        \hline
    \end{tabular}

    \vskip 3mm
    
    \begin{tabular}{ccccc}
        \hline
        \multirow{2}{2.5cm}{\centering Transonic (Sec. \ref{sec:foil})} & 
        \multirow{2}{2.0cm}{\centering Size} &
        \multirow{2}{2.0cm}{\centering Loss} &
        \multirow{2}{1.5cm}{\centering Epoch ratio} &
        \multirow{2}{2.5cm}{\centering Parameter ratio}
        \\ 
        
        \\
        \hline
        \noalign{\vskip 1mm}
        
        Classical&  40-4& $\sim10^{-2}$&  0.310& 1.000\\
        Quantum&  2& $\sim10^{\ 1}$&  $>$ 1& 0.018\\
        Hybrid&  40-4-2& $\sim10^{-1}$&  0.725&  0.509\\
        
        \noalign{\vskip 1mm}
        \hline
    \end{tabular}

    \caption{Benchmarked performance of classical, quantum and hybrid PINN models in terms of loss, epoch ratio and parameter ratios (definitions in text), based on representive model size for each architecture. }
    \label{tab:summary}
\end{table}

Key findings:
\begin{enumerate}
    \item \textbf{Quantum PINNs excel in harmonic regimes}, offering high accuracy with very low parameter costs (Table \ref{tab:summary}) due to the Fourier structure of PQCs (Section~\ref{sec:smooth}). However, quantum models have low trainability and struggle to generalize to discontinuities or shocks (Sections~\ref{sec:shock},~\ref{sec:foil}).
    
    \item \textbf{HQPINNs can mitigate spurious artifacts}, including overshoot near discontinuities (Fig.~\ref{fig:fig5}), by leveraging classical sub-networks to handle non-harmonic features. 
    
    \item \textbf{HQPINNs can match the best performance of classical models} even for non-harmonic problems (Fig.~\ref{fig:fig6}d), at reduced parameter costs. This suggests that, in certain cases, the quantum sub-network could complement the training of models with non-harmonic features.
    
    \item \textbf{HQPINNs are highly trainable} for both harmonic and non-harmonic problems as shown by the respective epoch ratios (Table \ref{tab:summary}). Note however that this trainability is reduced if the quantum layer were under-parameterized (Fig.~\ref{fig:fig8}).
    
    \item \textbf{Quantum emulation cost is a constraint}. Simulating PQCs classically results in per-epoch runtimes up to 100× slower than fully classical models (Table~\ref{tab:foil}, Appendix \ref{sec:tables}); scalability is a practical concern on current hardware.
\end{enumerate}

Overall, HQPINNs can potentially offer competitive performance at economical parameter costs across both harmonic and non-harmonic problem types, without falling decisively short in any single performance metric (Table~\ref{tab:summary}).

\section{Conclusion}\label{sec:conclusion}

Here we briefly review and discuss the implementation of a HQPINN model for high-speed flow problems, compared to classical and quantum PINN models. 

For problems with harmonic solutions (Sec. \ref{sec:smooth}), quantum models easily achieve high solution accuracies using very few trainable parameters but with low trainability. For problems with non-harmonic solutions (Sec. \ref{sec:shock}), classical models are significantly more accurate and trainable than quantum models but this also comes at a significant parameter cost. Either way, hybrid models are balanced in all aspects of accuracy, trainability and efficiency (Table \ref{tab:summary}). 

In conclusion, the HQPINN is a highly dependable, problem agnostic architecture that can adapt useful features from both classical and quantum sub-networks. Based on our benchmarked study, we find that the HQPINN is currently limited by the relatively small scale of quantum sub-network which is unable to cope with more complex shock features of the aerofoil problem, which degrades the overall performance. Nevertheless, with rapid progress in quantum technologies, larger scale quantum hardware will eventually be realized and we believe hybrid PINN architectures will eventually find use in broad applications, especially where the nature of the solution is not known \emph{a-priori}. 

In practice, a hybrid classical-quantum architecture can be implemented in a CPU-QPU environment, leveraging the capabilities of the latest quantum chips and processors in the pipeline, such as Google's Willow \cite{Acharya2024} and Microsoft's Majorana I \cite{Aghaee2025}. Realistically, the quantum processing unit (QPU) can efficiently process the quantum layer for harmonic components, while the CPU handles the classical layer for non-harmonic components. Future work may include adaptive weight balancing between classical and quantum sub-network layers, use of GPU-accelerated classical simulation of quantum models and fault-tolerant quantum hardware as they become available.

\section*{Acknowledgements}
This research is supported by the National Research Foundation, Singapore and the Agency for Science, Technology and Research (A*STAR) under Quantum Engineering Programme (NRF2021-QEP2-02-P03), Hybrid Quantum-Classical Computing
HQCC 1.0 (SC23/24-7534CI) and A*STAR C230917003.

\section*{Author Declarations}
The authors have no conflicts to disclose.

\section*{Data Availability}
The data that support the findings of this study are available from the corresponding author upon reasonable request. 

\bibliographystyle{unsrt}
\bibliography{bib}

\clearpage
\appendix

\renewcommand\thefigure{\thesection\arabic{figure}}
\setcounter{figure}{0} 
\renewcommand{\thetable}{A\arabic{table}} 
\setcounter{table}{0} 

\section{\label{sec:Appendix}Appendix}

\subsection{\label{sec:circuit}Parameterized quantum circuit design}

The performance of QML depends on the expressivity and trainability of the model~\cite{Abbas2021}, and even within the simple design framework of alternating feature map and ansatz layers, there are many configurations to choose from \cite{Siegl2024}. 

\begin{figure}[ht]
\centering
\[ 
\Qcircuit @C=0.6em @R=.6em {
 &  & & \mbox{$A$} 
& &  & & & & & & \mbox{$S$} & & & & &  &  
\\~\\
& \lstick{\ket{0}}\qw & \gate{R_Z(\theta_{1,1})} & \gate{R_X(\theta_{2,1})} & \gate{R_Z(\theta_{3,1})} & \ctrl{1}   & \qw & \qw &  \targ & \qw & \qw& \gate{R_Y(\varphi_1)} & \qw & \qw
\\ 
& \lstick{\ket{0}}\qw & \gate{R_Z(\theta_{1,2})} & \gate{R_X(\theta_{2,2})} & \gate{R_Z(\theta_{3,2})} & \targ & \ctrl{1} & \qw & \qw & \qw & \qw& \gate{R_Y(\varphi_2)} & \qw & \qw
\\ 
& \lstick{\ket{0}}\qw & \gate{R_Z(\theta_{1,3})} & \gate{R_X(\theta_{2,3})} & \gate{R_Z(\theta_{3,3})} & \qw & \targ & \ctrl{1} & \qw & \qw & \qw& \gate{R_Y(\varphi_3)} & \qw & \qw
\\ 
& \lstick{\ket{0}}\qw & \gate{R_Z(\theta_{1,4})} & \gate{R_X(\theta_{2,4})} & \gate{R_Z(\theta_{3,4})} & \qw & \qw & \targ & \ctrl{-3} & \qw & \qw& \gate{R_Y(\varphi_4)} & \qw & \qw 
\gategroup{3}{3}{6}{9}{.6em}{.}
\gategroup{3}{11}{6}{12}{.6em}{.}
}
\]
\caption{Example of an ansatz layer $A(\theta)$ consisting of $RZ-RX-RZ$ parameterized gate combination and $CNOT$ entanglers, and a feature map layer $S(\varphi)$ using $RY$ parameterized gates to perform angle encoding repeated for data re-uploading. $\theta$ and $\varphi$ refer to parameter weights and input features respectively.}
\label{fig:figA1}
\end{figure}
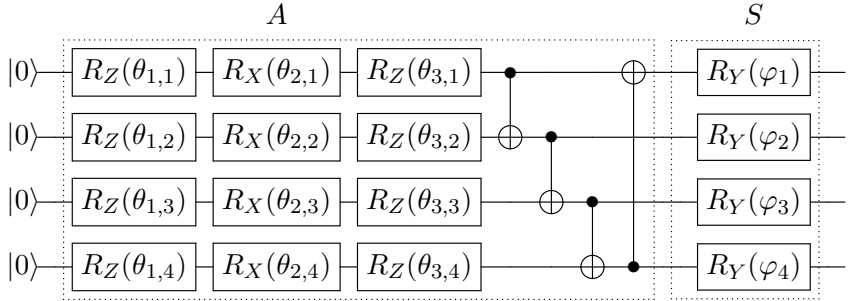

Following previous works\cite{Kordzanganeh2023, xiao2024physics, Siegl2024}, we employ a four-qubit PQC as an entangled $RZ-RX-RZ$ gates for an ansatz layer $A(\theta)$, and a $RY$ gates for a feature map layer $S(\varphi)$ (Fig. \ref{fig:figA1}). In this study, we consider a PQC with the following gate sequence: $A-(S-A)_1- \dots-(S-A)_{l-1}$. 

\subsection{\label{sec:harmonic_oscillator}Damped harmonic oscillator}

The quantum Fourier model (Eq.~\ref{eq:Truncated Fourier}) could extract harmonic features from a suitable dataset \cite{Kordzanganeh2023} and solve PDEs with harmonic features \cite{xiao2024physics, Siegl2024}. Here, we test the use of a hybrid classical-quantum model on a simple problem of a 1D damped harmonic oscillator, which is also an introductory PINN problem \cite{PINN_intro}. The displacement $u$ in time $t$ is described by a 1D second-order ordinary differential equation (ODE) as, 
    \begin{align}
    m\partial_{tt} u + \mu\partial_t u +ku = 0, \quad t \in (0,1],
    \label{eq:damped oscillator}
\end{align}
where $m$ is the mass of the oscillator, $\mu$ is the coefficient of friction and $k$ is the spring constant. The PINN loss function is
    \begin{align}
    \mathcal{L} = (u_{NN}(0)-1)^2 + \lambda_1 \left( \partial_t u_{NN}\right)^2+\frac{\lambda_2}{N}\sum_i^N \left( \left[ m \partial_{tt} + \mu \partial_{t} + k \right] u_{NN}(t_{i}) \right)^2,
    \label{eq:oscillator loss}
\end{align}
where $u_{NN}(t)$ is neural network solution. 

Following~\cite{PINN_intro}, we set $m=1$, $\mu=4$, $k=400$, $\lambda_1 = 10^{-1}$ and $\lambda_2 = 10^{-4}$ and implement PINN on \emph{Pytorch}~\cite{Paszke2019} and \emph{Pennylane}~\cite{Bergholm2018} using Adam optimizer with initial learning rate 0.002 up to $2000$ epochs. The parallel neural network layers in this test are combinations of the following: 
\begin{description}
    \item[Quantum layer] PQC with 3 qubits and 3 layers. 
    \item[Classical layer] Multi-layer perceptron with 2 hidden layers with 16 nodes each.
\end{description}

\begin{figure}
\centering
\includegraphics[width=0.9\linewidth]{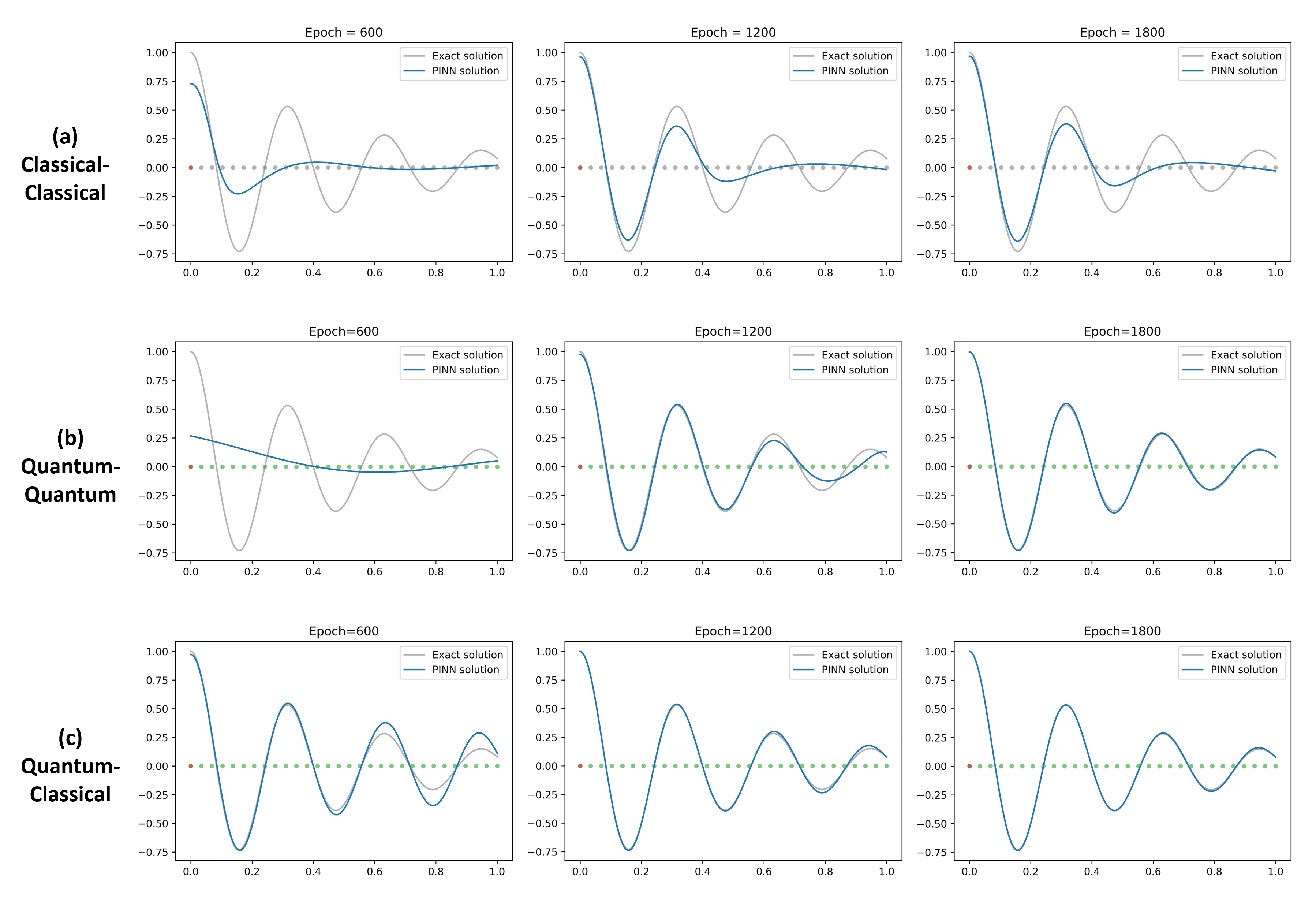}
\caption{Plots of $u_{NN}$ for $t \in (0,1]$  (blue curves) compared to the exact harmonic solution (grey curves) at training times 600, 1200 and 1800 (left to right). (a) The classical--classical network responds quickly but converges poorly. (b) The quantum--quantum network (middle) responds slowly but converges strongly due to the quantum Fourier model. (c) The hybrid quantum--classical network combines training features from classical and quantum layers and is able to converge quickly and with high fidelity.}
\label{fig:figA2}
\end{figure}

As proof of concept, we tested three parallel HQPINN configurations, namely classical--classical, quantum--quantum and hybrid quantum--classical neural networks~\cite{Kordzanganeh2023}. Figure~\ref{fig:figA2} shows that classical--classical networks struggles to learn the damped oscillations and requires more iterations to achieve convergence ($\approx15,000$ epochs~\cite{PINN_intro}). The quantum--quantum network converges with high fidelity, demonstrating the power of the quantum Fourier model in modeling harmonic solutions~\cite{schuld2021effect,xiao2024physics,Siegl2024}. Notably however, the quantum model is worse than classical model at early data fitting. By combining complementary features from classical and quantum layers, the hybrid model outperforms classical and quantum neural networks (Fig.~\ref{fig:figA2}c).

\subsection{\label{sec:cfd} Computational Fluid Dynamics simulation}

Computational Fluid Dynamics (CFD) simulation of steady-state transonic flow (Eq. \ref{eq:Euler} and \ref{eq:EOS}) past a NACA2012 airfoil is implemented on Ansys Fluent 2022 R2 \cite{ansys_fluent} using pressure-based solver on inviscid pressure-velocity coupled scheme. The inviscid flow assumption retains key features such as shock waves, and is commonly employed in PINN studies \cite{Mao2020, Liu2024}. 

For spatial discretization, we apply the least squares cell-based discretization for gradients, second-order discretization for pressure, and second-order upwind discretization for density, momentum, and energy. Residual of $10^{-6}$ is set as the convergence criterion for all variables.

\begin{figure}
\centering
\includegraphics[width=1\linewidth]{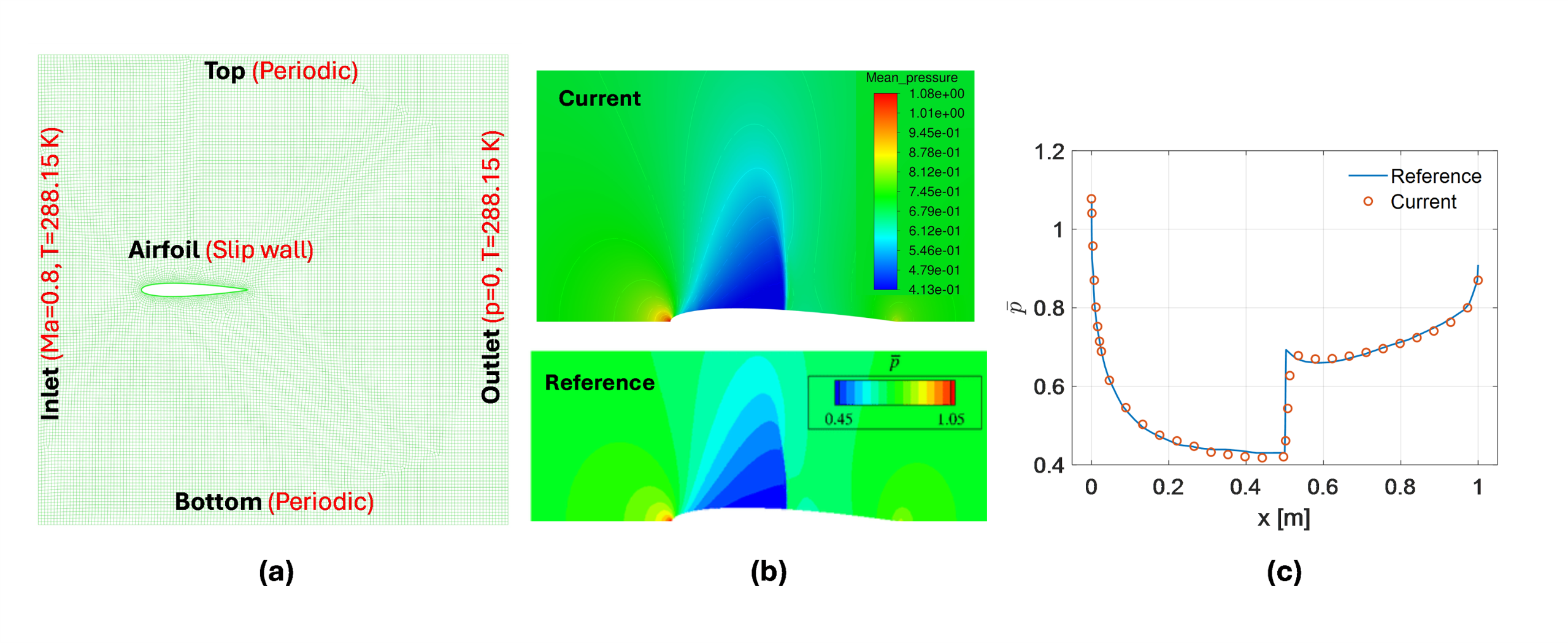}
\caption{(a) Computational domain for modeling a transonic flow past a NACA0012 airfoil with 1 m span in a 4.5 m by 4.5 m square grid. Compare (b) normalized mean pressure field over airfoil and (c) normalized pressure along surface chord, between current simulation and reference \cite{Gill2016}.}
\label{fig:figA3}
\end{figure}

Figure \ref{fig:figA3} shows the computational domain for a 2D compressible flow over a NACA0012 airfoil with 1 m span in a 4.5 m by 4.5 m square grid. Periodic boundary conditions are applied to both the top and bottom of the domain. We specify an inlet airflow velocity at Mach number 0.8 and a temperature 288.15 K. The outlet boundary is set at null air pressure and a temperature of 288.15 K. Free slip condition is applied at the surface of the airfoil. The finite volume grid consists of 25\,000 cells with mesh refinement so that cell widths are approximately 30 mm in the domain and 10 mm near the airfoil surface. Mesh convergence study is conducted.

Figure \ref{fig:figA3}(b) shows normalized mean pressure field over airfoil for current simulation and the reference \cite{Gill2016}. The far-field subsonic flow at Mach 0.8 transitions to transonic behavior as it passes over the NACA0012 airfoil, creating a normal shock near the trailing edge. Figure \ref{fig:figA3}(c) compares normalized pressure spanning the airfoil chord. Our results from the current simulation exhibit good agreement with reported values \cite{Gill2016}.

\subsection{\label{sec:quantum_hybrid} Quantum and hybrid models for transonic flow}

\begin{figure}[H]
\centering
\includegraphics[width=1\linewidth]{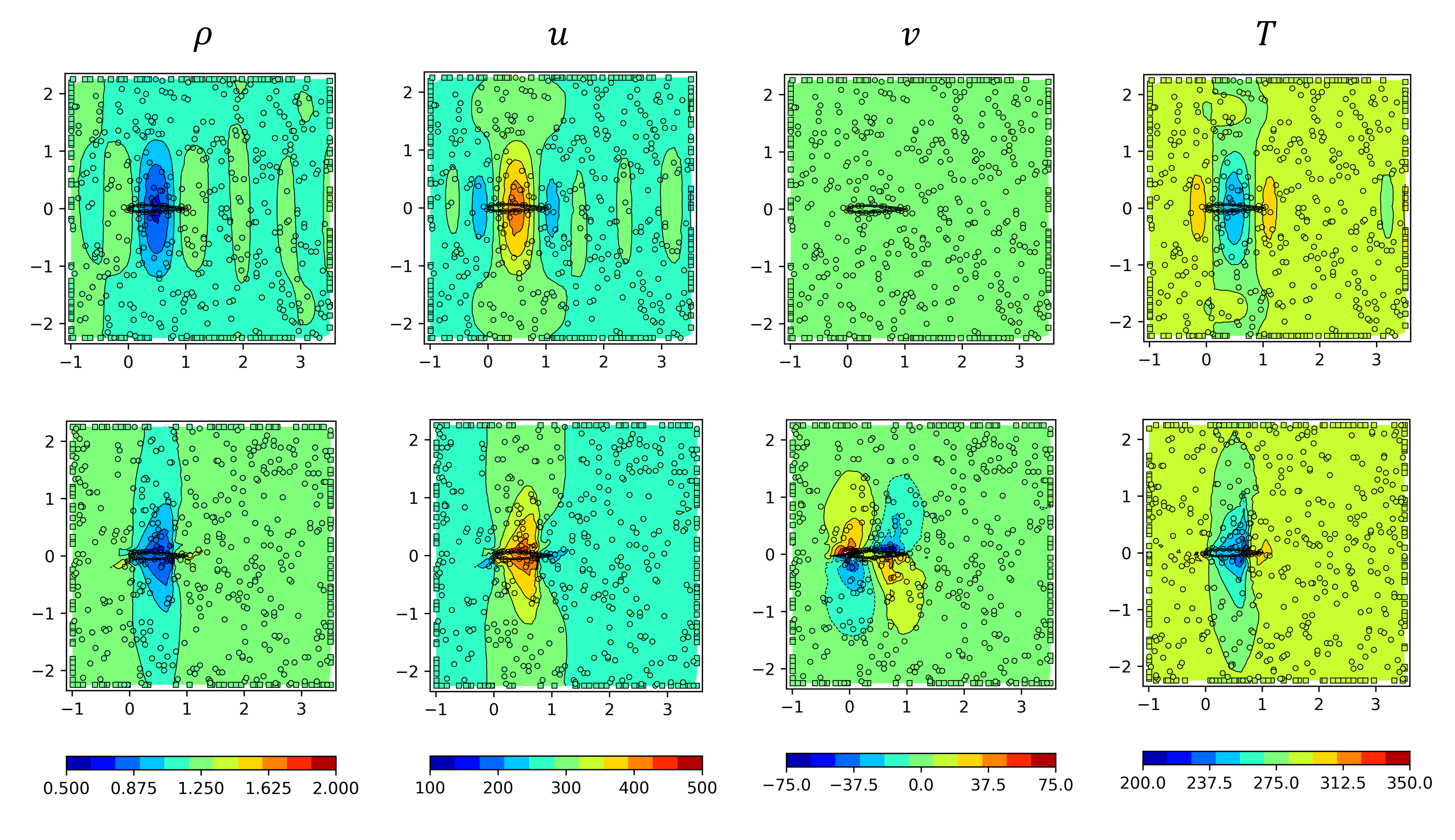}
\caption{Solution plots of density ($\rho$), horizontal and vertical velocities ($u,v$), and temperature ($T$) for (top) quantum \emph{qq-2} and (bottom) hybrid \emph{hy-40-4-2} models. Training points in open circles.}
\label{fig:figA4}
\end{figure}

Here we examine sample solutions for 2D transonic flow problem (Sec. \ref{sec:foil}) for quantum and hybrid models (Fig. \ref{fig:figA4}). The quantum model strives to fit a harmonic solution over discontinuous data training points and completely neglects vertical velocity solution due to under-parameterization. The hybrid model captures the characteristic transonic shock features using the classical network (compare Fig. \ref{fig:fig7}).

\subsection{\label{sec:tables} Tables for performance benchmarking}

\begin{table}[H]
    \centering
    \begin{tabular}{ccccccc}
        \hline
        \noalign{\vskip 2mm}  
        \multirow{2}{1cm}{\centering Model}&  \multirow{2}{1cm}{\centering Size}& \multirow{2}{2cm}{\centering Trainable parameters} & \multirow{2}{2cm}{\centering Loss} & \multirow{2}{2cm}{\centering Density error} & \multirow{2}{2cm}{\centering Pressure error} \\
        \\
        \noalign{\vskip 1mm}
        \hline
        \noalign{\vskip 2mm}
        
        \multirow{3}{1cm}{\centering cc} 
        &  10--4&  845&  $6.13\times10^{-7}$& $1.59\times10^{-6}$& $3.28\times10^{-9}$& \\
        &  10--7& 1505&  $\mathbf{3.16\times10^{-7}}$& $7.33\times10^{-7}$ & $2.22\times10^{-9}$& \\
        &  20--4& 2845&  $6.65\times10^{-7}$& $2.02\times10^{-6}$ & $3.96\times10^{-9}$& \\
        \noalign{\vskip 2mm}
        
        \multirow{2}{1cm}{\centering qq} 
        & 2  & 87(+36) & $\mathbf{2.63\times10^{-8}}$& $9.76\times10^{-9}$& $1.01\times10^{-9}$& \\
        & 3  & 87(+54) & $1.71\times10^{-7}$& $1.34\times10^{-7}$& $1.87\times10^{-9}$& \\     
        & 4  & 87(+72) & $2.75\times10^{-7}$& $2.80\times10^{-7}$& $2.46\times10^{-9}$& \\
        \noalign{\vskip 2mm}  
        
        \multirow{3}{1cm}{\centering hy} 
        & 10--4--2 & 448(+36) & $\mathbf{1.98\times10^{-7}}$& $2.29\times10^{-7}$& $8.44\times10^{-9}$& \\
        & 10--7--2 & 778(+36) & $3.31\times10^{-7}$&$3.24\times10^{-7}$& $9.04\times10^{-9}$& \\
        & 20--4--2 &1448(+36) & $8.62\times10^{-7}$& $3.19\times10^{-7}$& $4.98\times10^{-9}$& \\
        \noalign{\vskip 2mm}  

        \hline
    \end{tabular}
    \caption{Smooth Euler Equation (Section \ref{sec:smooth}). Trainable parameter count excludes quantum parameter count in parenthesis. Mean loss and relative error for density and pressure at the 20\,000\ts{th} epoch, averaged over 6 runs. Best loss results are shown in bold.}
    \label{tab:smooth}
\end{table}

\begin{table}[H]
    \centering
    \begin{tabular}{ccccccc}
        \hline
        \noalign{\vskip 2mm}  
        \multirow{2}{1cm}{\centering Model}&  \multirow{2}{1cm}{\centering Size}& \multirow{2}{2cm}{\centering Trainable parameters} & \multirow{2}{2cm}{\centering Loss} & \multirow{2}{2cm}{\centering Density error} & \multirow{2}{2cm}{\centering Pressure error} \\
        \\
        \noalign{\vskip 1mm}
        \hline
        \noalign{\vskip 2mm}
        
        \multirow{3}{1cm}{\centering cc} 
        &  10--4&  845&  $\mathbf{6.89\times10^{-7}}$& $2.89\times10^{-4}$& $1.92\times10^{-6}$& \\
        &  10--7& 1505&  $1.39\times10^{-6}$& $4.07\times10^{-4}$ & $6.06\times10^{-7}$& \\
        &  20--4& 2845&  $1.51\times10^{-6}$& $4.67\times10^{-4}$ & $6.59\times10^{-7}$& \\
        \noalign{\vskip 2mm}
        
        \multirow{2}{1cm}{\centering qq} 
        & 2  & 87(+36) & $5.00\times10^{-4}$& $1.20\times10^{-3}$& $4.51\times10^{-5}$& \\
        & 3  & 87(+54) & $1.45\times10^{-4}$& $1.30\times10^{-3}$& $1.53\times10^{-5}$& \\     
        & 4  & 87(+72) & $\mathbf{1.39\times10^{-4}}$& $9.65\times10^{-4}$& $1.24\times10^{-5}$& \\
        \noalign{\vskip 2mm}  
        
        \multirow{3}{1cm}{\centering hy} 
        & 10--4--2 & 448(+36) & $7.53\times10^{-6}$& $5.29\times10^{-4}$& $1.96\times10^{-6}$& \\
        & 10--7--2 & 778(+36) & $3.93\times10^{-5}$& $3.55\times10^{-4}$& $3.52\times10^{-5}$& \\
        & 20--4--2 &1448(+36) & $\mathbf{6.98\times10^{-6}}$& $2.37\times10^{-4}$& $3.29\times10^{-6}$& \\
        \noalign{\vskip 2mm}  

        \hline
    \end{tabular}
    \caption{Discontinuous Euler Equation (Section \ref{sec:shock}). Trainable parameter count excludes quantum parameter count in parenthesis. Mean loss and relative error for density and pressure at the 20\,000\ts{th} epoch, averaged over 6 runs. Best loss results are shown in bold.}
    \label{tab:shock}
\end{table}

\begin{table}[H]
    \centering
    \begin{tabular}{ccccccc}
        \hline
        \noalign{\vskip 2mm}  
        \multirow{2}{1cm}{\centering Model}&  \multirow{2}{1cm}{\centering Size}& \multirow{2}{2cm}{\centering Trainable parameters} & \multirow{2}{2cm}{\centering Time (s/epoch)} & \multirow{2}{2cm}{\centering Loss} & \multirow{2}{2cm}{\centering Pressure error} \\
        \\
        \noalign{\vskip 1mm}
        \hline
        \noalign{\vskip 2mm}

        \multirow{3}{1cm}{\centering cc} 
        &  40--4&  10\,628&  0.11& $1.11\times10^{-2}$& $1.35\times10^{-3}$& \\
        &  40--7&  20\,468&  0.21& $\mathbf{2.48\times10^{-3}}$& $6.79\times10^{-4}$& \\
        &  80--4&  40\,388&  0.20& $3.13\times10^{-3}$& $7.74\times10^{-4}$& \\
        \noalign{\vskip 2mm}

        \multirow{3}{1cm}{\centering qq} 
        &  2&  140(+48)&  1.9& $9.67\times10^{1}$& $3.59\times10^{-3}$& \\
        &  4&  140(+96)&  4.6& $8.48\times10^{1}$& $2.53\times10^{-3}$& \\
        &  6&  140(+144)&  9.0& $\mathbf{6.65\times10^{1}}$& $4.21\times10^{-3}$& \\
        \noalign{\vskip 2mm}

        \multirow{3}{1cm}{\centering hy} 
        & 40--4--2&  5360(+48)&  1.3& $1.36\times10^{-1}$& $4.05\times10^{-3}$& \\
        & 40--7--2&  10\,280(+48)&  1.1& $3.71\times10^{0}$& $1.95\times10^{-3}$& \\
        & 80--4--2&  20\,240(+48)&  1.1& $\mathbf{2.56\times10^{-2}}$& $2.46\times10^{-3}$& \\
        \noalign{\vskip 2mm}
        
        \hline
    \end{tabular}
    \caption{2D Transonic Aerofoil Flow (Section \ref{sec:foil}). Trainable parameter count excludes quantum parameter count in parenthesis. Average wall time shown in seconds per epoch. Mean loss and relative error for pressure averaged over 6 runs at the 42\,000\ts{th} epoch. Best loss results are shown in bold.}
    \label{tab:foil}
\end{table}
\end{document}